\documentstyle[12pt]{article}
\def\sqr#1{\mathop{\mkern0.5\thinmuskip
\vbox{\hrule\hbox{\vrule
\hskip#1\vrule height#1 width 0pt\vrule}\hrule}
\mkern0.5\thinmuskip}}
\def\Square{\mathchoice{\sqr{6pt}}{\sqr{6pt}}
{\sqr{4pt}}{\sqr{3pt}}}
\def\sq{\Square}  
\def\II{{\rm 1\!\!I}}

\def\tr{{\rm tr\,}}

\def\log{{\rm log\,}}

\def\be{\begin{equation}}
\def\ee{\end{equation}}
\def\bea{\begin{eqnarray}}
\def\eea{\end{eqnarray}}

\begin{document}

\begin{titlepage}

\null
\vskip-3truecm
\hspace*{7truecm}{\hrulefill}
\par
\vskip-4truemm
\par
\hspace*{7truecm}{\hrulefill}
\par
\vskip5mm
\par
\hspace*{7truecm}{{\large\sf University of Greifswald (February, 1997)}}
\vskip4mm
\par
\hspace*{7truecm}{\hrulefill}
\par
\vskip-4truemm
\par
\hspace*{7truecm}{\hrulefill}
\par
\bigskip
\hspace*{7truecm}hep-th/9703005
\bigskip
\par
\hspace*{7truecm}{\sf submitted to:}\par
\hspace*{7truecm}{\sf }\par
\hspace*{7truecm}{\large\sf PHYSICS LETTERS B}\par
\hspace*{7truecm}{\sf }\par
\smallskip

\vskip2truecm

\centerline{\LARGE\bf Singularities of Green functions of the}
\medskip
\centerline{\LARGE\bf products of the Laplace type operators}
\bigskip

\bigskip
\bigskip
\centerline{\Large\bf I. G. Avramidi
\footnote{On leave of absence from Research Institute for Physics, 
Rostov State University,  Stachki 194, 344104 Rostov-on-Don, Russia.}
\footnote{\sc E-mail: avramidi@rz.uni-greifswald.de}}
\bigskip

\centerline{\it Department of Mathematics, University of Greifswald}
\centerline{\it F.-L.-Jahnstr. 15a, D--17489 Greifswald, Germany}

\bigskip
\centerline{\today}
\medskip

\vfill

{\narrower
\par
The structure of diagonal singularities of Green functions of partial differential operators 
of even order acting on smooth sections of a vector bundle over a Riemannian manifold is studied.
A special class of operators formed by the products of second-order operators 
of Laplace type defined with the help of a  unique
Riemannian metric and a unique bundle connection but with 
different potential terms is investigated.
Explicit simple formulas for singularities of Green functions of such operators
in terms of the usual heat kernel coefficients
are obtained.
\par}
\end{titlepage}


\baselineskip=18pt
\section{Introduction}

The Green functions $G(x,x')$ of a partial differential operator,
$L:$ $C^\infty(V(M))$ $\to $ $C^\infty(V(M))$, 
acting on fields $\varphi\in C^\infty(V(M))$, i.e. smooth 
sections of  a vector bundle $V(M)$, 
over a $d$-dimensional Riemannian manifold $M$ with metric $g$
is of great importance in mathematical physics and quantum theory
\cite{hadamard23,dewitt65,gilkey84,barv-v85a}.
In particular, the singularities of the Green functions on the diagonal
$x=x'$ play a crucial role in the renormalization procedure of quantum field theory
\cite{bogolyubov76}.

In this paper we consider differential operators of higher orders of a special
form, namely, the products of  the second-order operators of Laplace type, i.e.,
\be
L=F_N\cdots F_2F_1,
\label{1a}
\ee
where 
\be
F_i=-\Square+Q_i , \qquad i=1,2,\dots,N,
\ee
with $\Square=g^{\mu\nu}\nabla_\mu\nabla_\nu$ being the generalized Laplacian 
(or D'Alambertian, in hyperbolic case), $\nabla$ the covariant derivative
on the vector bundle $V(M)$, $Q_i=Q_i(x)$ are arbitrary matrix-valued functions, 
i.e. endomorphisms of the vector bundle $V(M)$, called the potential terms.
Operators of this type arise in higher-derivative field theories  (for example, in
higher-derivative quantum gravity
\cite{avr86b}).
We do not restrict the signature of the metric of the manifold, it can be Euclidean,
${\rm sign}g=(+\cdots+)$, leading to {\it elliptic} operators, as well as Minkowskian 
${\rm sign}g=(-+\cdots+)$, corresponding to {\it hyperbolic} ones.

The Green functions of the operator $L$ are defined by requring them to satisfy the
equation
\be
LG(x,x')=\delta(x,x'),
\label{1}
\ee
with $L$ acting on the first argument of $G(x,x')$ and 
$\delta(x,x')$ being the covariant delta-distribution on the vector bundle $V(M)$. 

Here some remarks are to be made:
\begin{itemize}
\item[i)]
First, in Euclidean (elliptic) case for noncomplete manifolds with boundary, the equation 
(\ref{1}) does not define a unique Green function and 
one has to impose some suitable {\it boundary conditions}. We assume that some
appropriate boundary conditions are choosen, 
so that the Green function is defined uniquely.
Since we are going to investigate only the singularities of $G(x,x')$ near diagonal
$x=x'$,
our analysis will be {\it purely local} and the boundary conditions will not play 
any role in our consideration.

\item[ii)]
Second, in pseudo-Euclidean (hyperbolic) case we will fix the Green function by
the {\it Wick rotation}, i.e. the analytic continuation  from the Euclidean case, which
is equivalent to adding an infinitesimal negative imaginary part to all potential terms,
$Q_j\to Q_j-i\varepsilon$. In other words, we consider the {\it Feynman propagators}. 
Thus in the following we deal mainly with the elliptic case---the formulas
for the hyperbolic one are obtained just by analytic continuation.

\item[iii)]
Last, we will also assume for convenince the potential terms to be 
everywhere large enough, $Q_i(x)>{\rm const}>0$,
so that all operators $F_i$ are {\it elliptic} and {\it  positive}. 
Then the operator $L$ is, of course, also positive.
The singularities,
do not depend, of course, on this assumption but it 
simplifies the intermediate computations.

\end{itemize}

\section{Heat kernel representation of the Green function}

The Green function of a product of operators (\ref{1a}) is obviously
expressed as the convolution of the Green functions $G_j=F^{-1}_j$ 
of each factor
\be
G=L^{-1}=F_1^{-1}F^{-1}_2\cdots F_N^{-1}=G_1 G_2\cdots G_N.
\ee
Further, by using the heat kernel representation for the positive
elliptic Laplace type operators
\cite{dewitt65,gilkey84,avr86b,avr91b}
\be
G_j=F_j^{-1}=\int\limits_0^\infty dt_j U_j(t_j),
\ee
where 
\be
U_j(t_j)\equiv \exp({-t_jF_j}),
\ee
one obtains the heat kernel representation for the Green function $G$
\be
G=\int\limits_0^\infty dt_1\cdots \int\limits_0^\infty dt_N U_1(t_1)\cdots U_N(t_N),
\label{2a}
\ee
or in the kernel form
\bea
G(x,x')&=&\int\limits_0^\infty dt_1\cdots \int\limits_0^\infty dt_N 
\int\limits_M d{\rm vol}(y_1)\cdots \int\limits_M d{\rm vol}(y_{N-1})
U_1(t_1|x,y_1)
\nonumber\\[10pt]
&&
\times U_2(t_2|y_1,y_2)\cdots U_N(t_N|y_{N-1},x'),
\label{3}
\eea
where $d{\rm vol}(y)=d y g^{1/2}(y)$ is the invariant volume element on $M$.

The behavior of the Green function $G(x,x')$ near diagonal $x\to x'$ 
is determined by the behavior of the heat kernels $U_j(t_j|y_{j-1},y_j)$ 
at $t_j\to 0$, which is well known
\cite{gilkey84,dewitt65,avr86b,avr91b}.
Thus, following
\cite{avr91b}, we use a semi-classical ansatz 
for the heat kernel
\be
U_j(t_j|y_{j-1},y_j)=(4\pi)^{-d/2}t^{-d/2}\Delta^{1/2}(y_{j-1},y_j)
\exp\left(-{\sigma(y_{j-1},y_j)\over 2t_j}\right)\Omega_j(t_j|y_{j-1},y_j).
\label{6aa}
\ee
Here $\sigma(y_{j-1},y_j)$ is the geodetic interval defined as one half the square 
of the geodesic distance between points $y_{j-1}$ and $y_j$ and $\Delta(y_{j-1},y_j)$
is the corresponding Van Fleck determinant
\cite{dewitt65,dewitt60,synge60}.
This ansatz reproduces the initial condition for the heat kernel,
\be
\lim_{t_j\to 0}U_j(t_j|y_{j-1},y_j)=\delta(y_{j-1},y_j),
\ee
provided 
the function $\Omega_j(t_j|y_{j-1},y_j)$, called the transfer function, 
has a regular limit as $t\to 0$
\cite{avr86b,avr91b}
\be
\lim_{t_j\to 0}\Omega_j(t_j|y_{j-1},y_j)={\cal P}(y_{j-1},y_j),
\ee
where ${\cal P}(y_{j-1},y_j)$ is the parallel displacement operator along the geodesic.

Now let us  introduce new integration variables
\be
t=t_1+\cdots+t_N, \qquad \xi_j={t_j\over t},
\ee
with the variables $\xi_j$ satisfying a constraint
\be
\xi_1+\cdots+\xi_N=1.
\ee
The integration over the $t_j$'s takes the form
\be
\int\limits_0^\infty dt_1\cdots \int\limits_0^\infty dt_N\, f(t_j)
=\int\limits_0^\infty dt\,{t^N\over (N-1)!}
\int\limits_0^1 d\xi_1\cdots \int\limits_0^1 d\xi_N
\delta(\xi_1+\cdots+\xi_N-1)\, f(t\xi_j)
\label{4a}
\ee
and, therefore, the Green function (\ref{3}) can be written as
\be
G(x,x')=(4\pi)^{-d/2}{\Delta^{1/2}(x,x')\over (N-1)!}
\int\limits_0^\infty dt\,t^{N-d/2-1}
\exp\left(-{\sigma(x,x')\over 2t}\right)A(t|x,x'),
\label{5}
\ee
where
\bea
A(t|x,x')&=&(4\pi)^{-(N-1)d/2}
\int\limits_M d{\rm vol}(y_1)\cdots \int\limits_M d{\rm vol}(y_{N-1})
{\Delta^{1/2}(x,y_1)\cdots\Delta^{1/2}(y_{N-1},x')\over\Delta^{1/2}(x,x')}
\nonumber\\
&&
\times\int\limits_0^\infty d\xi_1\cdots \int\limits_0^\infty d\xi_N
\delta(\xi_1+\cdots+\xi_N-1)\,
(\xi_1\cdots\xi_N)^{-d/2}
\nonumber\\
&&
\times\exp\left(-{1\over 2t}\Phi(x,y_1,\dots,x';\xi_1,\xi_N)\right)
\Omega_1(t\xi_1|x,y_1)
\nonumber\\[10pt]
&&
\times\Omega_2(t\xi_2|y_1,y_2)
\cdots\Omega_N(t\xi_N|y_{N-1},x'),
\label{6}
\eea
where
\be
\Phi(x,y_1,\dots,x';\xi_1,\xi_N)
={\sigma(x,y_1)\over \xi_1}
+{\sigma(y_1,y_2)\over \xi_2}
+{\sigma(y_{N-1},x')\over \xi_N}
-\sigma(x,x').
\label{6b}
\ee
Remember, that locally, near diagonal, i.e. for $x$ close to $x'$, the Van Fleck determinant does not vanish
$\Delta(x,x')\ne 0$.

\section{Asymptotic expansions}

From the heat kernel representation (\ref{5}) it is clear that the diagonal singularities
of the Green function, when $\sigma(x,x')\to 0$, are determined by the behavior of the
function $A(t|x,x')$ at small $t$. To study this behavior we remember, 
first of all, the well known
asymptotic expansion of the transfer function 
near diagonal $y_{j-1}\to y_j$ as $t\to 0$
\cite{dewitt65,avr86b,avr91b}
\be
\Omega_j(t_j|y_{j-1},y_j)\sim 
\sum\limits_{k\ge 0}{(-t_j)^k\over k!}
a^{(j)}_k(y_{j-1},y_j),
\label{6a}
\ee
where $a^{(j)}_k(y_{j-1},y_j)$ are the famous 
Hadamard-Minakshisundaram-De Witt-Seeley (HMDS)-coefficients
\cite{hadamard23,minakshisundaram53,dewitt65,seeley67b}.
They can be computed in form of covariant Taylor series
and are known in general case up to $a_4$
\cite{gilkey75b,avr86b,avr90b,avr91b,amsterdamski89}.
In particular cases, as in flat space, there results for higher-order
coefficients. For a review of the methods for calculation of HMDS-coefficients 
and further references see 
\cite{avr-s96,schimming93}.

By using the obvious equation
\be
{1\over \sqrt{4\pi t\xi_j}}
\exp\left(-{\sigma(y_{j_1},y_j)\over 2t\xi_j}\right)\Bigg|_{t\to 0}
=\delta(y_{j_1},y_j)
\ee
it is not difficult to show that the function $A(t|x,x')$ is regular in the limit $t\to 0$
\be
\lim_{t\to 0}A(t|x,x')={\cal P}(x,x').
\ee
Moreover, by changing the integration variables $y_j$ and using the
asymptotic expansion of the heat kernel (\ref{6a}) one can obtain the 
asymptotic expansion of the function $A(t|x,x')$, i.e.  as $t\to 0$ there holds, in particular,
\be
A(t|x,x')=\sum\limits_{k=0}^{\left[(d+1)/2\right]-N-1}
{(-t)^k\over k!}c_k(x,x')+R(t|x,x'),
\label{7}
\ee
where $c_k(x,x')$ are some coefficients which are analytic in $x\to x'$ and
the rest term is of order 
\be
R(t|x,x')\sim O\left(t^{\left[{(d+1)/2}\right]-N}\right).
\ee
The positivity of the operator $L$ leads, in addition, to exponential falling-off
of the function $A(t|x,x')$ at the infinity, as $t\to\infty$.
Thus the integral over $t$ in eq. (\ref{5}) always converges at the upper limit $t\to\infty$.

Using eq.  (\ref{7}) one can calculate the integral (\ref{5}) over $t$ 
and obtain an expansion of the Green function near diagonal, 
when $\sigma(x,x')\to 0$,
\bea
G(x,x')&=&(4\pi)^{-d/2}\sum\limits_{k=0}^{\left[{d+1\over 2}\right]-N-1}
(-1)^k{\Gamma({d\over 2}-N-k)\over k!(N-1)!}\Delta^{1/2}(x,x')
c_{k}(x,x')\left({2\over\sigma(x,x')}\right)^{{d\over 2}-N-k}
\nonumber\\[10pt]
&&
+\Psi(x,x'),
\eea
where
\be
\Psi(x,x')=(4\pi)^{-d/2}{\Delta^{1/2}(x,x')\over (N-1)!}
\int\limits_0^\infty dt\,t^{N-d/2-1}
\exp\left(-{\sigma(x,x')\over 2t}\right)R(t|x,x').
\label{9}
\ee

The first term in this formula is a polynomial in the inverse powers of $\sigma(x,x')$ 
and gives the leading singularities of the Green function.
Let us look at the second term, the function $\Psi(x,x')$.
Off the diagonal, i.e. for $x\ne x'$, or $\sigma\ne 0$, the integral over $t$ always 
converges at the limit $t\to 0$. However, on the diagonal, $\sigma=0$, the point is a bit more subtle. 
For {\it odd dimensions}, $d$, we have
\be
t^{N-d/2-1}R(t|x,x')\sim O(t^{-1/2})
\ee
and, therefore, the integral (\ref{9}) over $t$ converges at the limit $t\to 0$ also
on diagonal $x=x'$, meaning that the function $\Psi(x,x')$
is a regular smooth function on the diagonal, actually it is analytic in  
$x$ close to $x'$ and, hence, can be expanded in a covariant Taylor expansion.

For {\it even dimensions}, $d$, in contrary, there holds
\be
t^{N-d/2-1}R(t|x,x')\sim O(t^{-1}),
\ee
and, therefore, the integral (\ref{9}) does not converge on the diagonal,
for $\sigma= 0$. It is easy to see, that in this case the function $\Psi(x,x')$
exhibits a logarithmic divergence
\be
\Psi(x,x')={(4\pi)^{-d/2} (-1)^{d/2-N+1}\over (N-1)!
\Gamma({d\over 2}-N+1)}
 \Delta^{1/2}(x,x') c_{{d\over 2}-N}(x,x')\log\left({\sigma(x,x')\over 2\mu^2}\right)
+\Psi_{\rm reg}(x,x';\mu),
\label{8}
\ee
where $\Psi_{\rm reg}(x,x';\mu)$ is a regular analytic function in $x\to x'$ and
$\mu$ is an arbitrary dimensionful parameter introduced to preserve dimensions.

Thus we obtained the complete structure of the diagonal singularities of the 
Green function of the product of $N$ Laplace type operators 
\bea
G(x,x')&=&(4\pi)^{-d/2}\sum\limits_{k=0}^{\left[{d+1\over 2}\right]-N-1}
(-1)^k{\Gamma({d\over 2}-N-k)\over k!(N-1)!}\Delta^{1/2}(x,x')
c_{k}(x,x')\left({2\over\sigma(x,x')}\right)^{{d\over 2}-N-k}
\nonumber\\[10pt]
&&
+(4\pi)^{-d/2}{\Delta^{1/2}(x,x')
\over (N-1)!\Gamma({d\over 2}-N+1)}
\chi(x,x') \log{\sigma(x,x')\over 2\mu^2}
+G_{\rm reg}(x,x';\mu),
\label{8a}
\eea
where
\be
\chi(x,x')=\left\{
\begin{array}{ll}
(-1)^{d/2-N+1}c_{{d\over 2}-N}(x,x'),& {\rm for \ even\ } d\\
0,& {\rm for \ odd\ } d
\end{array}
\right.
\label{29a}
\ee
Therefrom, we see that:
\begin{itemize}
\item[i)]
the structure of singularities is determined  by the coefficients $c_k(x,x')$ 
of the asymptotic expansion of the function $A(t|x,x')$;

\item[ii)]
in odd dimenions there is no logarithmic singularity and the regular part of the Green
function $G_{\rm reg}$ does not depend on $\mu$;

\item[iii)]
{}for $N>d/2$ there are {\it no} singularities at all and the Green function is regular
analytic function on the diagonal, i.e. there exists finite 
coincidence limit $G(x,x)$;

\item[iv)]
{}for $N=\left[(d+1)/2\right]$ there are no singularities for odd dimension 
$d=2N-1$ and there 
is only logarithmic singularity in even dimension $d=2N$.

\end{itemize}

The logarithmic singularity of the Green function is very important.
On the one hand it determines, as usual, 
the renormalization properties of the
regular part of the Green function. Indeed, 
noting that $G$ does not depend on $\mu$ we obtain from (\ref{8a})
\be
\mu{\partial\over\partial \mu}G_{\rm reg}
={2(4\pi)^{-d/2}\over (N-1)!\Gamma({d\over 2}-N+1)}
\Delta^{1/2}(x,x')\chi(x,x').
\ee

On the other hand, the absence of the 
logarithmic singularity is a necessary
condition for the validity
of the Huygence principle for hyperbolic operators
\cite{schimming78,schimming82,schimming90}.

The problem is now to calculate the coefficients $c_k(x,x')$.


\section{Particular cases}

\subsection{$N$-th power of a Laplace type operator}

The simplest case is, of course, the case of equal potential terms, i.e.
\be
Q_1=Q_2=\cdots=Q_N\equiv Q.
\ee
In this case all operators $F_j$ are equal to each other 
\be
F_1=\cdots=F_N\equiv F=-\sq+Q,
\ee
and the operator $L$ is the $N$-th power of the Laplace type operator
\be
L=F^N.
\ee
Using the heat kernel representation we find
\be
G=F^{-N}={1\over (N-1)!}\int\limits_0^\infty dt\, t^{N-1}U(t),
\ee
where $U(t)=\exp(-tF)$.
Rewriting this in kernel form and comparing with eq. (\ref{5}) we find 
that the function $A(t|x,x')$ in this case is just the transfer function determined 
by the ansatz (\ref{6aa}) for the heat kernel of the Laplace type operator $F$:
\be
A(t|x,x')=\Omega(t|x,x').
\label{10}
\ee
This means, in particular, that the transfer function satisfies a nontrivial identity 
\bea
\Omega(t|x,x')&=&(4\pi)^{-(N-1)d/2}
\int\limits_M d{\rm vol}(y_1)\cdots \int\limits_M d{\rm vol}(y_{N-1})
{\Delta^{1/2}(x,y_1)\cdots\Delta^{1/2}(y_{N-1},x')\over\Delta^{1/2}(x,x')}
\nonumber\\
&&
\times\int\limits_0^\infty d\xi_1\cdots \int\limits_0^\infty d\xi_N
\delta(\xi_1+\cdots+\xi_N-1)\,
(\xi_1\cdots\xi_N)^{-d/2}
\nonumber\\
&&
\times\exp\left(-{1\over 2t}\Phi(x,y_1,\dots,x';\xi_1,\xi_N)\right)
\Omega(t\xi_1|x,y_1)
\nonumber\\[10pt]
&&
\times\Omega(t\xi_2|y_1,y_2)\cdots\Omega(t\xi_N|y_{N-1},x'),
\label{6aaa}
\eea
where $\Phi(x,y_1,\dots,x';\xi_1,\xi_N)$ is defined by eq. (\ref{6b}).
{}From eq. (\ref{10}) we find that 
the coefficients $c_k$ are equal to the usual heat kernel coefficients
\be
c_k=a_k.
\ee

Thus the extent to which the coefficients $c_k$ differ from the HMDS-coefficients
$a_k^{(j)}$ in general case is only due to the different potential terms.
We see that the function $A(t)$ is an `averaged' (or weighted) transfer function of
the Laplace type operators.

\subsection{Covariantly constant perturbations of the potential terms}

Let us consider now the case when the potential terms $Q_j$ differ from each other by
covariantly constant terms $\lambda_j$ which commute with all other matrices 
(the potential terms, the bundle curvature ${\cal R}_{\mu\nu}=[\nabla_\mu,\nabla_\nu]$ 
and their derivatives), i.e.,
\be
Q_j\equiv Q+\lambda_j, \qquad \nabla_\mu \lambda_j=0, 
\ee
\be
[\lambda_j(x),Q(y)]=[\lambda_j(x),{\cal R}_{\mu\nu}(y)]
=[\lambda_i(x),\lambda_j(y)]=0.
\ee
In this case we have
\be
F_i=F+\lambda_i,
\ee
where
\be
F=-\sq+Q.
\label{11a}
\ee
and, therefore,
\be
L=(F+\lambda_N)\cdots(F+\lambda_2)(F+\lambda_1).
\ee
The operators $F_i$, even if different, commute
and from the heat kernel representation (\ref{2a}) 
by using the change of variables (\ref{4a})
we obtain 
\be
G=L^{-1}={1\over (N-1)!}\int\limits_0^\infty dt\,t^{N-1} 
\omega(t)\exp(-tF),
\label{12}
\ee
where
\be
\omega(t)=e^{-t\lambda_N}\cdot
{e^{-t(\lambda_1-\lambda_N)}-1\over t(\lambda_1-\lambda_N)}
\cdot
{e^{-t(\lambda_2-\lambda_N)}-1\over t(\lambda_2-\lambda_N)}
\cdots
{e^{-t(\lambda_{N-1}-\lambda_N)}
-1\over t(\lambda_{N-1}-\lambda_N)}
\ee
Rewriting this equation in the kernel form and comparing with eq. (\ref{5}) 
we find that the function $A(t)$ differs from the transfer
function $\Omega(t|x,x')$ for the Laplace type
operator $F$ (\ref{11a}) just by a prefactor $\omega(t)$:
\be
A(t|x,x')=\omega(t)\Omega(t|x,x').
\ee
The function $\omega(t)$ is analytic in $t$. 
Expanding it in the Taylor series
\be
\omega(t)=\sum_{n\ge 0}{(-t)^n\over n!}\omega_n
\ee
we find the coefficients $c_k$
\be
c_k=\sum_{0\le n\le k}{k\choose n}\omega_{n}a_{k-n},
\ee
where $a_n$ are the HMDS-coefficients.

\section{Algebraic approach}

As we have seen, the function $A(t|x,x')$ is very closely connected with the transfer function $\Omega(t|x,x')$ of an Laplace type operator.
Moreover, in particular cases considered above the coefficients $c_k$ are expressed {\it linearly} in terms of the usual HMDS-coefficients $a_k$.
Let us show now how one can compute the coefficients $c_k$ in general case.

Let us define some auxilliary operators $Z_j$, ($j=1,2,\dots,N$), by
\bea
Z_1&=&F_1G=F_2^{-1}\cdots F_N^{-1}\nonumber\\
Z_2&=&F_2Z_1=F_2F_1G=F^{-1}_3\cdots F_N^{-1}\nonumber\\
&\vdots&\nonumber\\
Z_{N-1}&=&F_{N-1}Z_{N-2}=F^{-1}_N\nonumber\\
Z_{N}&=&F_{N}Z_{N-1}=1
\label{13a}
\eea

Now let us introduce a column-vector built from the auxillary operators 
\be
Z=\left(
\begin{array}{c}
G\\
Z_1\\
Z_2\\
\vdots\\
Z_{N-1}
\end{array}
\right).
\ee
The Green function we are looking for is the first component of this vector and can be obtained by 
multiplying with the row-vector
$\Pi^{\dag}_1=(1,0,\dots,0)$ 
\be
G=\Pi^{\dag}_1 Z.
\label{14a}
\ee
Further, let us introduce a matrix $\widetilde F$ formed by the operators $F_i$
\be
\widetilde F=\left(
\begin{array}{cccccc}
F_1&-1&0&0&\dots&0\\
0&F_2&-1&0&\dots&0\\
\vdots&\vdots&\vdots&\vdots&\ddots&\vdots\\
0&0&0&0&\dots&F_N
\end{array}
\right)
\ee
Then the equations (\ref{13a}) can be rewritten in a matrix form
\be
\widetilde F Z=\Pi_N,
\label{14}
\ee
where
\be
\Pi_N=\left(
\begin{array}{c}
0\\
0\\
\vdots\\
0\\
1
\end{array}
\right)
\ee
The operator $\widetilde F$ has the form
\be
\widetilde F=-\II\cdot\sq+\widetilde Q,
\label{16a}
\ee
where $\II$ is the unity $N\times N$ matrix and $\widetilde Q$ is defined by
\be
\widetilde Q=\left(
\begin{array}{cccccc}
Q_1&-1&0&0&\dots&0\\
0&Q_2&-1&0&\dots&0\\
\vdots&\vdots&\vdots&\vdots&\ddots&\vdots\\
0&0&0&0&\dots&Q_N
\end{array}
\right)
\label{17a}
\ee
It is of Laplace type and is, therefore,
nondegenerate, i.e. there exists a well defined Green function $\widetilde G=\widetilde F^{-1}$.
Thus from the eq. (\ref{14}) we have
\be
Z=\widetilde G\Pi_N.
\ee
Finally, from (\ref{14a}) we get the Green function 
\be
G=\Pi^{\dag}_1 Z
=\Pi^{\dag}_1\widetilde G\Pi_N
={\rm tr}\,(\Pi\widetilde G),
\label{16}
\ee
where $\Pi$ is a projector of the form
\be
\Pi=\Pi_N\otimes\Pi^{\dag}_1=\left(
\begin{array}{cccc}
0&0&\dots&0\\
\vdots&\vdots&\vdots&\vdots\\
0&0&\dots&0\\
1&0&\dots&0
\end{array}
\right)
\ee
and ${\rm `tr'}$ denotes the usual matrix trace. 
Note that this trace has nothing to do with the trace over the bundle indices which are left intact.

This shows that the problem is reduced to a Laplace type operator 
$\widetilde F$ with additional matrix structure.
Using the usual heat kernel representtion for the Green function of the Laplace type operator we have from eq. (\ref{16})
\be
G(x,x')=(4\pi)^{-d/2}\Delta^{1/2}(x,x')
\int\limits_0^\infty dt\,t^{-d/2}
\exp\left(-{\sigma(x,x')\over 2t}\right)
{\rm tr\,}\Pi\,\widetilde \Omega(t|x,x'),
\label{17}
\ee
where $\widetilde\Omega(t|x,x')$ is the transfer function of the operator
$\widetilde F$ (\ref{16a}).
It is not difficult to show that due to the presence of the projector $\Pi$ and the special 
form of the potential term $\widetilde Q$ (\ref{17a}) 
the asymptotic expansion of the function ${\rm tr}\, 
\Pi\widetilde\Omega(t)$ as $t\to 0$ begins at the order $t^{N-1}$. Comparing the eq. (\ref{17}) 
with eq. (\ref{5}) we find then the function $A(t|x,x')$ in terms of the transfer function $\widetilde\Omega(t|x,x')$
\be
A(t)=(N-1)!\,t^{-N+1}\,{\rm tr\,}\Pi\,\widetilde \Omega(t).
\ee
By using the expansions of the functions $A(t)$ (\ref{7}) and $\widetilde\Omega(t)$ (\ref{6a}) 
and equating the powers of $t$ we obtain finally the coefficients $c_k$ in terms of the HMDS-coefficients
\be
c_k(x,x')={(-1)^k\over {k+N-1\choose k}}{\rm tr}\, 
(\Pi\,\widetilde b_{k+N-1}(x,x')),
\ee
where $\widetilde b_n$ are the HMDS-coefficients of the Laplace type operator $\widetilde F$ (\ref{16a}) with the potential term $\widetilde Q$ (\ref{17a}).

This formula gives the coefficients of the asymptotic expansion of the function $A(t|x,x')$ in 
terms of well known heat kernel coefficients of the Laplace type operators and 
determines finally the structure of the diagonal singularities of the product of Laplace type operators (\ref{8a}).

{}From (\ref{29a})  we obtain, in particular, 
the function $\chi(x,x')$ determining the logarithmic singularity
\be
\chi(x,x')=\left\{
\begin{array}{ll}
-{(N-1)!\Gamma(d/2-N+1)\over\Gamma(d/2)}\tr\,(\Pi\widetilde b_{d/2-1}(x,x')),& {\rm for \ even\ } d\\
0,& {\rm for \ odd\ } d
\end{array}
\right.
\ee

\section*{Acknowledgements}

I am greatly indebted to Rainer Schimming for many stimulating
and fruitful discussions. 
The warm hospitality at the University of Greifswald is also gratefully acknowledged.
This work was supported by the Deutsche Forschungsgemeinschaft.



\begin{thebibliography}{999}

\bibitem{hadamard23} J. Hadamard,
{\it Lectures on Cauchy's Problem}, in: {\it Linear Partial Differential Equations}
(Yale U. P., New Haven, 1923)
        
\bibitem{dewitt65} B. S. De~Witt, 
{\it Dynamical Theory of Groups and Fields}, 
(New York: Gordon and Breach, 1965)

\bibitem{gilkey84} P. B. Gilkey,
{\it Invariance Theory, the Heat Equation and the  Atiyah - Singer
Index Theorem} (Publish or Perish, Wilmington, DE, USA, 1984)

\bibitem{barv-v85a} A. O. Barvinsky and G. A. Vilkovisky, 
Phys. Rep. C {\bf 119} (1985) 1

\bibitem{bogolyubov76} N. N. Bogolyubov and D. V. Shirkov, 
{\it Introduction to the
theory of quantized fields}, (Moscow: Nauka, 1976)

\bibitem{avr86b} I. G. Avramidi, 
{\it Covariant Methods for the Calculation of 
the Effective Action in Quantum Field Theory and Investigation of 
Higher-Derivative Quantum Gravity}, 
PhD thesis, Moscow State University (1986), UDK 530.12:531.51,
[in Russian]; Transl.: hep-th/9510140

\bibitem{avr91b} I. G. Avramidi, 
Nucl. Phys. B {\bf 355} (1991) 712

\bibitem{dewitt60} B. S. De Witt and R. W. Brehme, Ann. Physics {\bf 9}
(1960) 220

\bibitem{synge60} J. L. Synge, 
{\it Relativity: The general theory}, (Amsterdam: North-Holland, 1960).

\bibitem{minakshisundaram53} S. Minakshisundaram, 
J. Indian Math. Soc. {\bf 17} (1953) 158

\bibitem{seeley67b} R. T. Seeley, 
Proc. Symp. Pure Math. {\bf 10} (1967) 288

\bibitem{gilkey75b} P. B. Gilkey, 
J. Diff. Geom. {\bf 10} (1975) 601

\bibitem{avr90b} I. G. Avramidi, 
Phys. Lett. B {\bf 238} (1990) 92

\bibitem{amsterdamski89} P. Amsterdamski, A. L. Berkin and 
D. J. O'Connor,
Class. Quantum  Grav. {\bf 6} (1989) 1981

\bibitem{avr-s96} I. G. Avramidi and R. Schimming, 
{\it Algorithms for the calculation of the heat kernel coefficients},
in: {\it `Quantum Field Theory under the Influence of 
External Conditions'}, Ed. M. Bordag, Teubner-Texte zur Physik, Band 30,  
(Stuttgart: Teubner, 1996), p.~150

\bibitem{schimming93} R. Schimming, {\it Calculation of the heat kernel coefficients}, 
in: {\it Analysis, Geometry and Groups. A Riemann Legacy Volume}, 
Eds. H. M. Srivastava and Th. M. Rassias, (Hadronic Press, Palm Harbour, 1993), 
p. 627

\bibitem{schimming78} R. Schimming, Beitr. z. Analysis, {\bf 11} (1978) 45

\bibitem{schimming82} R. Schimming, Z. f. Analysis u. ihre Anw. 
{\bf 2} (1982) 71

\bibitem{schimming90} R. Schimming, Math. Nachr. {\bf 147} (1990) 217


\end{thebibliography}
\end{document}